\pdfoutput=1
\documentclass[sigconf,nonacm]{acmart}
\usepackage{needspace}
\usepackage{placeins}
\usepackage{xcolor}

\NewDocumentCommand{\yang}{ mO{} }{\textcolor{red}{\textsuperscript{\textit{yang}}\textsf{\textbf{\small[#1]}}}}

\newcommand{\sstab}{\rule{0pt}{8pt}\\[-2.2ex]}
\newcommand{\stitle}[1]{\sstab\noindent{\bf #1}}

\setcopyright{none}
\title{MIRA: Evidence-Verified Repair Memory for Text-to-SQL Correction}
\newcommand{\method}{\textsc{MIRA}}

\author{Yining Liu}
\affiliation{
  \institution{Beijing Institute of Technology}
  \city{Zhuhai}
  \country{China}
}
\email{lynnn@bit.edu.cn}

\author{Chenyu Yang}
\affiliation{
  \institution{The Hong Kong University of Science and Technology (Guangzhou)}
  \city{Guangzhou}
  \country{China}
}
\email{cyang662@connect.hkust-gz.edu.cn}

\author{Boyan Li}
\affiliation{
  \institution{The Hong Kong University of Science and Technology (Guangzhou)}
  \city{Guangzhou}
  \country{China}
}
\email{bli303@connect.hkust-gz.edu.cn}

\author{Rui Mao}
\affiliation{
  \institution{Shenzhen University}
  \city{Shenzhen}
  \country{China}
}
\email{mao@szu.edu.cn}

\author{Yuyu Luo}
\affiliation{
  \institution{The Hong Kong University of Science and Technology (Guangzhou)}
  \city{Guangzhou}
  \country{China}
}
\email{yuyuluo@hkust-gz.edu.cn}

\makeatletter
\renewcommand{\@mkauthors}{%
  \global\setbox\mktitle@bx=\vbox{%
    \unvbox\mktitle@bx
    \centering
    {\large
      Yining Liu\textsuperscript{1},
      Chenyu Yang\textsuperscript{2},
      Boyan Li\textsuperscript{2},
      Rui Mao\textsuperscript{3}, and
      Yuyu Luo\textsuperscript{2,*}\par}
    \vspace{3pt}
    {\small
      \textsuperscript{1}Beijing Institute of Technology, Zhuhai, China\par
      \textsuperscript{2}The Hong Kong University of Science and Technology (Guangzhou), Guangzhou, China\par
      \textsuperscript{3}Shenzhen University, Shenzhen, China\par
      \vspace{2pt}
      Email: \href{mailto:lynnn@bit.edu.cn}{lynnn@bit.edu.cn},
      \href{mailto:cyang662@connect.hkust-gz.edu.cn}{cyang662@connect.hkust-gz.edu.cn},
      \href{mailto:bli303@connect.hkust-gz.edu.cn}{bli303@connect.hkust-gz.edu.cn}\par
      \href{mailto:mao@szu.edu.cn}{mao@szu.edu.cn},
      \href{mailto:yuyuluo@hkust-gz.edu.cn}{yuyuluo@hkust-gz.edu.cn}\par}
    \medskip}}
\makeatother

\begin{document}

\begin{abstract}
Text-to-SQL agents still produce executable yet semantically incorrect SQL. A reliable SQL corrector must repair incorrect queries without corrupting correct ones. Confirmed corrections from the same database can be reused without parameter updates. Existing methods, however, often bundle multiple errors and their repairs into a single coarse-grained experience. Applying the entire experience can introduce irrelevant edits and turn an initially correct query into an incorrect one. Reliable reuse therefore depends on three decisions: what to retain from a historical correction, when to activate the resulting memory, and how to adapt it to the current SQL. We propose \method{} (\textbf{M}emory-\textbf{I}tem \textbf{R}euse and \textbf{A}daptation), a pluggable SQL corrector that uses database evidence to guide memory reuse. \method{} converts historical corrections into independently reusable repair memory items. For each current query, it retrieves memory items using the question and SQL. It then checks each item against database evidence and adapts the supported items to the current SQL. We evaluate 1,785 test queries generated by three Text-to-SQL agents across 14 databases from BIRD and ScienceBenchmark. \method{} improves execution accuracy by 16.53\% and 8.78\% on BIRD and ScienceBenchmark, respectively. 

\end{abstract}


\maketitle

\begingroup
\renewcommand{\thefootnote}{*}
\footnotetext{Corresponding author.}
\endgroup

\section{Introduction}
\label{sec:introduction}
Text-to-SQL systems translate natural-language questions into executable SQL, providing users with convenient access to structured data~\cite{kim2020naturallanguage}. Cross-domain benchmarks such as Spider and BIRD evaluate this capability over diverse schemas and databases~\cite{yu2018spider,li2023bird}. Yet generated SQL can still misinterpret question semantics, schema meaning, or database values; many such errors execute successfully but return incorrect results~\cite{zhong2020semantic}. Post-generation \emph{SQL correction} is therefore important for improving Text-to-SQL reliability. Given a natural-language question, a database, and a fixed first-attempt SQL query produced by an upstream agent, a corrector must decide whether and how to revise the query without access to the gold SQL. Because the first-attempt SQL may be either incorrect or already correct, reliable correction must repair errors without corrupting correct predictions.

Existing methods enhance SQL correction in three main ways. The first is \textbf{self-correction}, in which the same model inspects and revises its own SQL through decomposition, reflection, or repeated inference~\cite{pourreza2023dinsql,shen2024selectsql}. However, a model that failed to derive the correct query may also fail to identify its own error. A recent survey finds that self-correction without reliable external feedback rarely yields consistent gains~\cite{kamoi2024selfcorrection}. The second is \textbf{training-based correction}, which trains a dedicated model on correction data so that common errors and repairs are encoded in its parameters~\cite{chen2023sqlcorrection,qu2025share,hong2026errorllm}. These models can learn recurring repair patterns, but new databases or error types may require retraining or adaptation, increasing cost and potentially limiting generalization. The community has therefore begun to explore \textbf{experience-based correction}, which invokes external historical experience at inference time. A common approach asks a language model to summarize historical corrections and feedback into corrective experience that guides future revisions~\cite{askari2025magic}. \textbf{\textit{However, these experience units can be coarse grained and may conflate several errors or repair requirements.}} Injecting them into a reasoning model may therefore introduce noise, yielding limited gains or even causing regression on queries that were already correct.

\begin{figure}[t]
  \centering
  \includegraphics[width=\columnwidth]{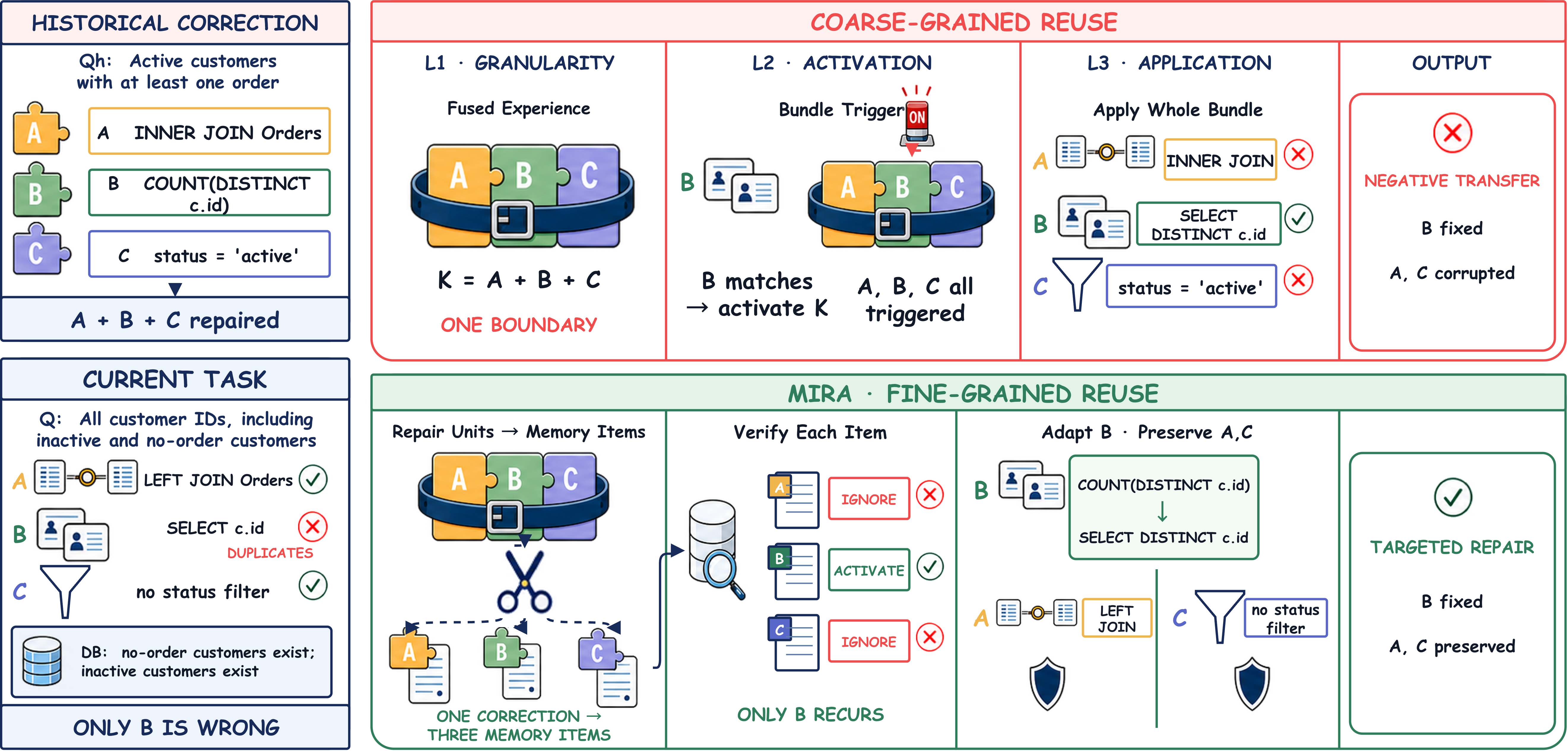}
  \caption{Motivation for \method{}. A historical correction fuses three repairs: requiring an order (A), deduplicating customers (B), and retaining only active customers (C). The current SQL violates only B. Coarse-grained reuse triggers all three and causes negative transfer, whereas \method{} separates the repairs, verifies them individually, and adapts only B while preserving A and C.}
  \Description{The top row shows a historical correction whose three repairs are fused into one coarse knowledge unit. Applying the unit to a current SQL that contains only duplicated customer IDs fixes that error but incorrectly replaces a left join with an inner join and adds an active-status filter. The bottom row shows MIRA separating the correction into three repair units, verifying that only the deduplication unit applies, adapting it from COUNT DISTINCT to SELECT DISTINCT, and preserving the current left join and absence of a status filter.}
  \label{fig:repair-unit-motivation}
  \vspace{-2em}
\end{figure}

\textbf{(L1)} A historical correction may contain several distinct errors and their repairs, but coarse-grained distillation can fuse them into one experience. The resulting corrective information cannot be reused independently. In Figure~\ref{fig:repair-unit-motivation}, the historical SQL simultaneously requires an order to exist (A), customer deduplication (B), and an active-customer restriction (C). The three repairs are bound into a single knowledge unit.

\textbf{(L2)} The trigger is defined over the entire fused experience and cannot judge each repair separately. Once a broad condition is satisfied, the complete experience may be activated. In Figure~\ref{fig:repair-unit-motivation}, the current SQL contains only the duplicated-customer error (B). Because it remains related to the historical case through customers and orders, however, the experience containing A, B, and C is activated as a whole. Database probing shows that customers without orders and inactive customers both exist. Repairs A and C must therefore be excluded rather than activated.

\textbf{(L3)} Once the entire experience is activated, coarse-grained application can transfer every source-case modification to the current SQL. In Figure~\ref{fig:repair-unit-motivation}, \texttt{DISTINCT} fixes the duplicated-customer error (B), but the transferred \texttt{INNER JOIN} (A) and \texttt{status='active'} predicate (C) remove customers without orders and inactive customers, causing negative transfer. A fine-grained method should instead adapt the historical \texttt{COUNT(DISTINCT c.id)} repair into the \texttt{SELECT DISTINCT c.id} required by the current task while preserving its correct \texttt{LEFT JOIN} and absence of a status filter.

Historical corrections are useful only if their repairs can be separated, verified, and adapted. Our goal is to reuse them without task-specific parameter updates, repairing recurring errors while preserving queries that are already correct. This goal raises three challenges.

\textbf{(C1: What to Remember) } A historical correction may fix several errors at once, so the resulting experience can mix distinct repairs. Without manual error annotations, the system must determine where one reusable repair ends and another begins. Each memory item must retain all changes needed for one repair, even when they span several SQL locations, while excluding unrelated changes

\textbf{(C2: When to Activate)} Semantic or structural relatedness indicates only that historical experience may be useful; it does not establish that the current SQL contains the same error. Without access to the gold SQL, the system must determine whether the targeted error actually occurs before allowing the memory to influence correction.

\textbf{(C3: How to Adapt)} Historical cases and current tasks commonly differ in context and SQL implementation, so source-case repairs cannot be copied directly. The system must translate the corrective requirement into changes that fit the current task while preserving logic that is already correct.

To address these challenges, we propose \method{} (\textbf{M}emory-\textbf{I}tem \textbf{R}euse and \textbf{A}daptation), which constructs independently reusable repair memory items from historical corrections, activates retrieved items using the current question, SQL, and database evidence, and adapts the supported items to the current SQL.

\textbf{To address C1,} \method{} introduces Repair Memory Construction. Starting from a historical incorrect SQL query and its confirmed correction, the framework first recovers a validated repair SQL anchored to the historical incorrect SQL. It then groups structural changes and uses the historical requirement, execution contrasts, and database facts to identify independently judgeable repair units. Each accepted repair unit becomes one memory item that records its semantic requirement, incorrect structure, target local database check, and source support. This design determines what should be remembered while excluding unrelated changes.

\textbf{To address C2,} \method{} first retrieves candidate items through semantic and structural matches to the current question and SQL. It then tests whether the error captured by each item recurs in the current SQL. An item is activated to guide correction only when the current task requirement, SQL form, and database evidence all support this recurrence.

\textbf{To address C3,} \method{} introduces Target Local Adaptation. For each activated memory item, the framework binds its repair requirement to the relevant tables, columns, values, and locations in the current SQL. It uses the item's preservation constraints to retain unrelated correct logic and jointly integrates multiple activated items into one complete rewrite. Historical corrections are therefore not copied verbatim but adapted to the current task.

\Needspace{6\baselineskip}
This paper makes the following contributions:

\textbf{1. A history-augmented SQL correction framework.} We propose \method{}, a memory-augmented framework grounded in database evidence that can serve as a post-processing corrector for different upstream Text-to-SQL agents. \method{} grounds memory item activation in the current question, SQL, and database evidence, then adapts the supported items to the current SQL, reducing regressions caused by memory misuse and cross-case transfer.

\textbf{2. Repair Memory Construction at independently judgeable repair boundaries.} We organize historical memory around repair units rather than complete correction cases. The method separates structurally coherent change groups and uses historical requirements and database evidence to retain independently judgeable repair units, providing a reusable basis for retrieval, activation, and adaptation to the current task.

\textbf{3. An empirical evaluation across upstream agents and databases.} We evaluate first-attempt SQL generated by CHESS~\cite{talaei2024chess}, DeepEye-SQL~\cite{li2025deepeyesql}, and OmniSQL-32B~\cite{li2025omnisql} on 14 databases from BIRD and ScienceBenchmark. \method{} improves execution accuracy by 16.53 and 8.78 percentage points on the two benchmarks, respectively, repairs 261 initially incorrect SQL queries, and causes regressions on only 18 initially correct queries.

\section{Problem Setting}
\label{sec:problem-setting}

\subsection{Post-generation SQL Correction}
\label{sec:sql-correction}

We study SQL correction as a post-generation task for Text-to-SQL. Let \(q\) denote a natural-language question together with any task evidence available to the corrector, and let \(\mathcal D\) denote a relational database, including its schema and data instance. Given \(q\), \(\mathcal D\), and the current SQL \(s^{\mathrm{cur}}\) produced by an upstream Text-to-SQL system, a corrector outputs an enhanced SQL \(\widetilde{s}\). We treat the upstream generator as a black box and observe only its output SQL. The current SQL may already be correct, so the corrector must not turn it into an incorrect query.

Let \(s^{*}\) be the gold SQL for \(q\). We regard a SQL query as correct when its execution result is equivalent to that of \(s^{*}\) on \(\mathcal D\):
\[
  \operatorname{Exec}_{\mathcal D}\!\left(\widetilde{s}\right)
  \equiv_{\mathrm{exec}}
  \operatorname{Exec}_{\mathcal D}\!\left(s^{*}\right).
\]
Here, \(\equiv_{\mathrm{exec}}\) denotes the execution-result comparison used in evaluation~\cite{zhong2020semantic}. The gold SQL is used only for evaluation and is unavailable during correction. The corrector must therefore rely on \(q\), \(s^{\mathrm{cur}}\), \(\mathcal D\), and the permitted historical information.

\subsection{SQL Correction with Historical Corrections}
\label{sec:historical-experience-setting}

As a Text-to-SQL system serves the same database over time, incorrect SQL queries and their confirmed corrections accumulate~\cite{urban2025feedback}. These records show which errors occurred and how they were corrected. However, one correction can combine several repairs with database-specific or incidental changes. The problem is to determine which repairs should become reusable repair memory items, whether their recorded errors recur, and how supported items should be adapted to the current SQL.

For database \(\mathcal D\), let the available historical corrections be
\[
  \mathcal H_{\mathcal D}
  =
  \left\{
    h_i
  \right\}_{i=1}^{N},
  \qquad
  h_i
  =
  \left(
    q_i,\,
    s_i^{-},\,
    s_i^{+}
  \right),
\]
where \(q_i\) is a historical task, \(s_i^{-}\) is its historical incorrect SQL, and \(s_i^{+}\) is its confirmed corrected SQL. All tuples concern the same database and share its schema and data instance. The complete set is available before a future correction task.

Given \(\mathcal H_{\mathcal D}\), a history-augmented SQL corrector \(F\) receives a current task \(q\), its current SQL \(s^{\mathrm{cur}}\), and database \(\mathcal D\), but not the gold SQL \(s^{*}\). It returns
\[
  \widetilde{s}
  =
  F\!\left(
    q,\,
    s^{\mathrm{cur}},\,
    \mathcal D,\,
    \mathcal H_{\mathcal D}
  \right).
\]
The historical corrections provide auxiliary information rather than gold supervision for the current task.

We focus on history-augmented correctors that derive database-scoped repair memory items from \(\mathcal H_{\mathcal D}\) before future tasks. At inference time, a memory item may guide correction only when the current task, SQL, and database evidence support recurrence of its recorded error. An unsupported item must not affect the output.

For a finite evaluation set of \(M\) scorable tasks, the change in execution accuracy is
\[
  \Delta\mathrm{EX}
  =
  \frac{N_{\mathrm{fix}}-N_{\mathrm{reg}}}{M},
\]
where \(N_{\mathrm{fix}}\) counts initially incorrect queries corrected successfully and \(N_{\mathrm{reg}}\) counts initially correct queries changed to incorrect outputs. Our objective is to increase successful repairs while keeping regressions low.

\section{Method}
\label{sec:method}

\subsection{Overview}
\label{sec:method-overview}

\begin{figure*}[t]
  \centering
  \includegraphics[width=\textwidth]{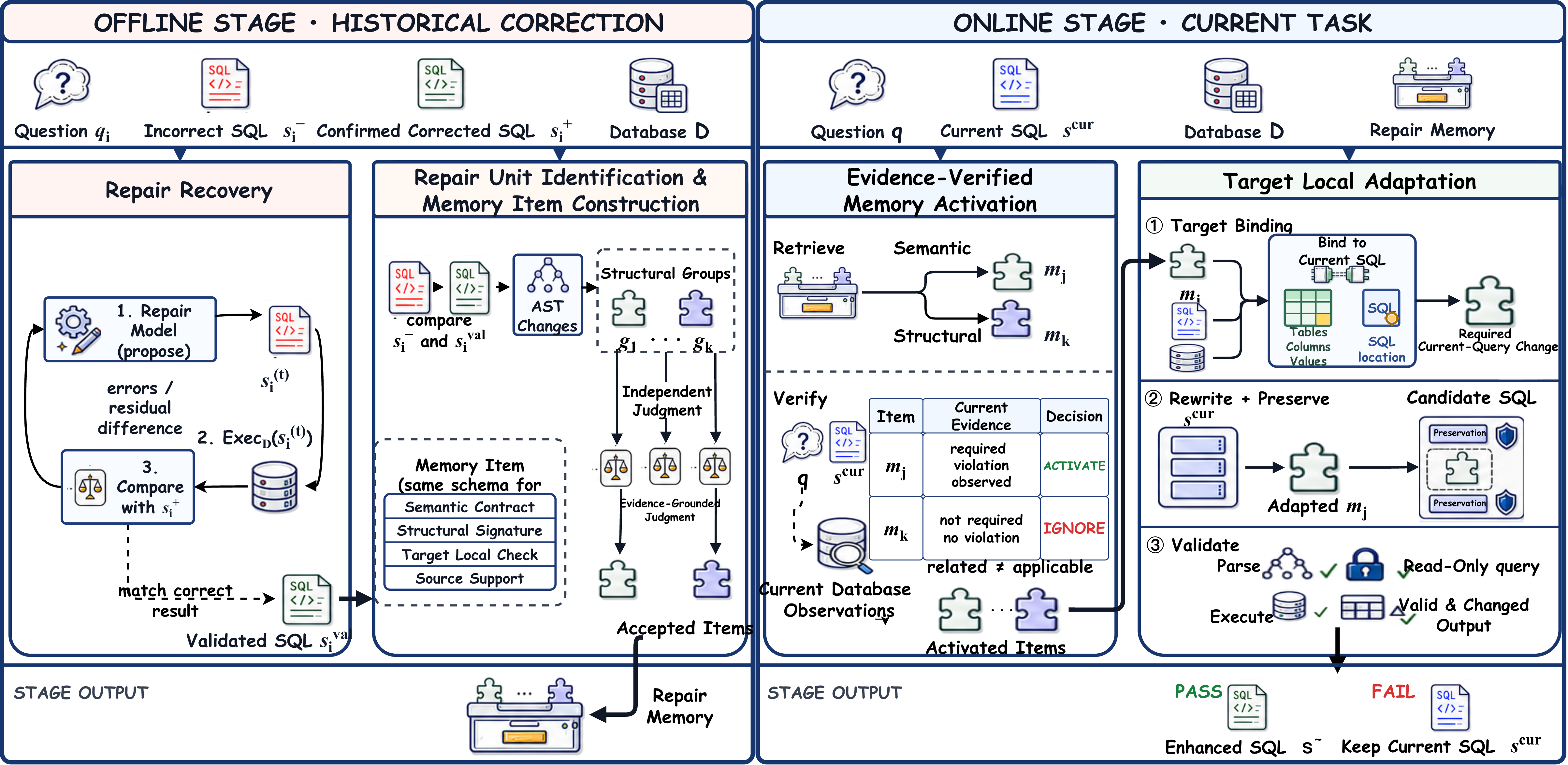}
  \caption{Overview of \method{}. Repair Memory Construction converts historical corrections from one database into a fixed repair memory. For a current task, Repair Memory Reuse and Adaptation retrieves candidate memory items, verifies whether their recorded errors recur, and adapts the activated items to the current SQL. The candidate rewrite is returned only if it passes mechanical validation; otherwise, \method{} retains the current SQL.}
  \Description{The offline stage receives a historical question, incorrect SQL, confirmed corrected SQL, and database. It recovers a validated repair SQL, separates its changes into repair units, judges each unit, and stores the accepted units as memory items in a database-scoped repair memory. The online stage receives a current question, current SQL, database, and repair memory. It retrieves memory items through semantic and structural channels, verifies them using current database observations, binds activated items to the current SQL, and validates the resulting candidate. A valid candidate becomes the enhanced SQL; otherwise, the current SQL is retained.}
  \label{fig:method-overview}
\end{figure*}

Figure~\ref{fig:method-overview} summarizes the two stages of \method{}. Repair Memory Construction processes confirmed corrections from one database and builds a fixed, database-scoped repair memory. Repair Memory Reuse and Adaptation uses this memory to correct future SQL queries.

Offline, \method{} receives a historical correction \((q_i,s_i^{-},s_i^{+})\) and database \(\mathcal D\). The confirmed corrected SQL \(s_i^{+}\) is not necessarily a minimal repair of \(s_i^{-}\), and one correction may contain several repairs. \method{} therefore starts from \(s_i^{-}\) to recover a validated repair SQL, separates its changes into independently judged repair units, and stores each accepted unit as one memory item. Section~\ref{sec:repair-memory-construction} describes this stage.

Online, \method{} receives a current task \(q\), current SQL \(s^{\mathrm{cur}}\), database \(\mathcal D\), and the corresponding repair memory. Retrieval proposes semantically or structurally related memory items, but relatedness alone does not show that their recorded errors recur. \method{} verifies each candidate against the current task, SQL, and database observations. It then binds the activated repair requirements to the current SQL, preserves unrelated logic, and returns the rewrite only if it passes mechanical validation. Otherwise, it retains \(s^{\mathrm{cur}}\). Section~\ref{sec:repair-memory-reuse} describes this stage.

\subsection{Offline Stage: Repair Memory Construction}
\label{sec:repair-memory-construction}

Repair Memory Construction must recover reusable repairs from a historical correction without manual error annotations. A historical correction can contain several SQL changes. Some may be unrelated to the errors being corrected, e.g., alias renaming. A single correction can also repair multiple errors, and each repair may require one or more structural changes. Direct SQL differencing therefore mixes incidental changes with repairs and does not reveal where one repair ends and another begins. \method{} addresses this problem in three steps: Repair Recovery produces a validated repair SQL from the historical incorrect SQL; Repair Unit Identification separates and verifies its repair requirements; and Memory Item Construction stores each accepted repair unit as one memory item.

\subsubsection{Repair Recovery}
\label{sec:repair-recovery}

For a historical correction \(h_i=(q_i,s_i^{-},s_i^{+})\) on database \(\mathcal D\), \(s_i^{+}\) provides a reference answer but is not necessarily a minimal repair of \(s_i^{-}\). It may rename aliases, reorder equivalent operations, or rewrite logic unrelated to the error. A direct diff between \(s_i^{-}\) and \(s_i^{+}\) therefore mixes required repairs with incidental changes. A one-shot model proposal may also remain invalid or leave residual result differences.
Generate-and-validate program repair similarly searches candidate changes and judges them with behavioral tests~\cite{legoues2012genprog,legoues2019repair}; here, the offline reference result supplies the validation target.

\stitle{Propose, execute, and compare.}
At round \(t\), the repair model starts from \(s_i^{-}\) and produces a complete candidate \(s_i^{(t)}\). It receives the historical task \(q_i\), the database schema and facts, \(s_i^{+}\) as an offline reference, and feedback from earlier rounds when available. \method{} executes the candidate on \(\mathcal D\) and compares its result with that of \(s_i^{+}\). Execution errors or residual result differences are returned in the next round.
Execution outcomes have also been used as feedback signals for language-model tool use~\cite{qiao2024execution}.

The first candidate whose execution result matches that of \(s_i^{+}\) becomes the validated repair SQL \(s_i^{\mathrm{val}}\):
\[
  \operatorname{Exec}_{\mathcal D}\!\left(s_i^{\mathrm{val}}\right)
  \equiv_{\mathrm{exec}}
  \operatorname{Exec}_{\mathcal D}\!\left(s_i^{+}\right).
\]
This condition establishes only execution-result equivalence on \(\mathcal D\)~\cite{zhong2020semantic}. Unlike formal SQL-equivalence techniques that reason under declared query semantics~\cite{chu2017hottsql,zhou2022spes}, it does not prove equivalence over all database instances. The validated repair SQL is used to identify repair units; it is not stored as a memory item. These units come from the structural difference between \(s_i^{-}\) and \(s_i^{\mathrm{val}}\), not from the model's account of its edits. If \(s_i^{-}\) is not executable, already matches the reference result, or no candidate matches within the repair budget, the historical correction produces no memory item.

\begin{figure}[t]
  \centering
  \includegraphics[width=\columnwidth]{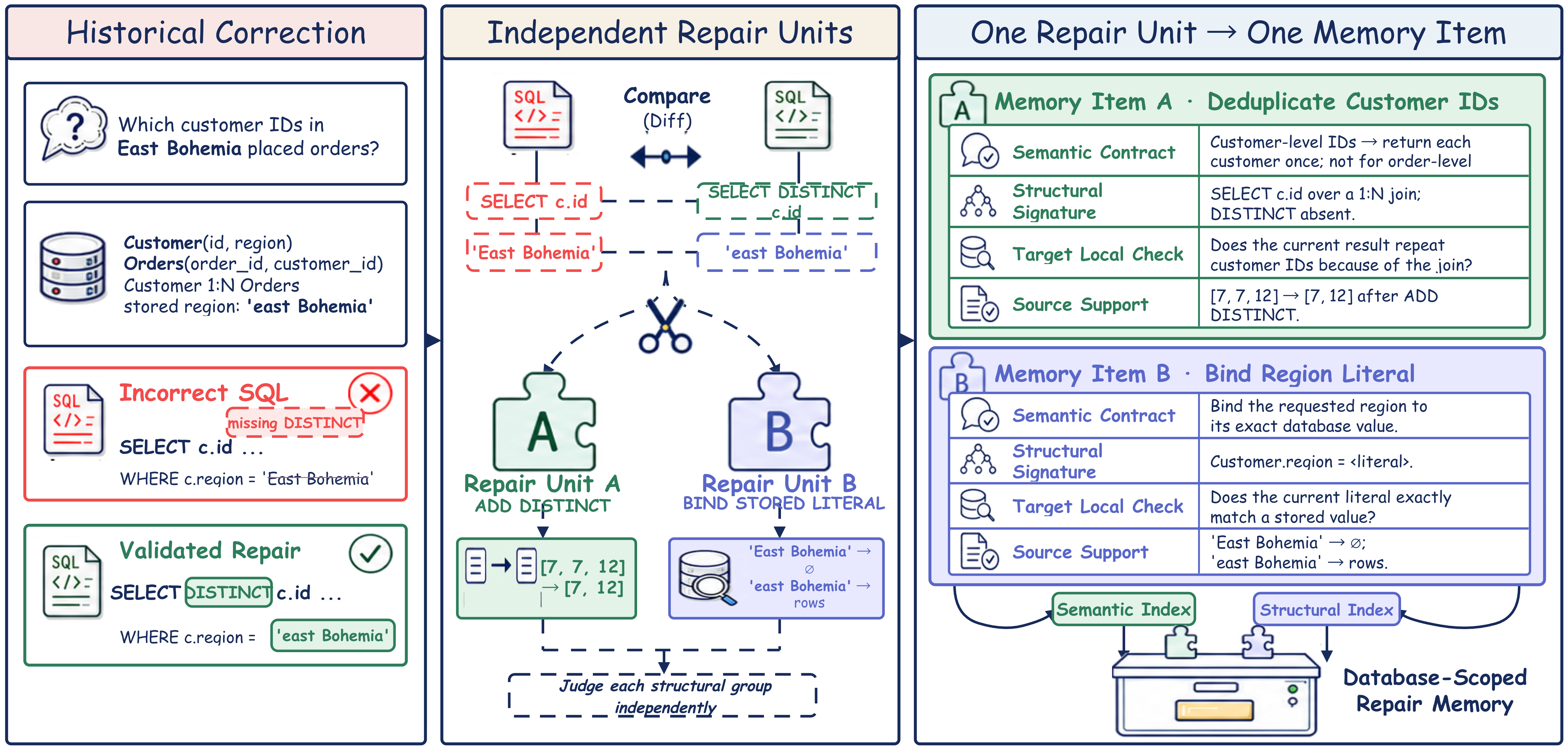}
  \caption{Repair memory construction for one historical correction. After Repair Recovery obtains \(s_i^{\mathrm{val}}\), \method{} groups the structural changes from \(s_i^{-}\) to \(s_i^{\mathrm{val}}\), judges each group independently, and stores each accepted repair unit as one memory item.}
  \Description{The left panel shows a historical question, database, incorrect SQL, and validated repair SQL. The middle panel compares the two SQL queries, separates their changes into a deduplication group and a database-literal binding group, and judges each structural group independently. The right panel converts each accepted repair unit into a memory item containing a semantic contract, structural signature, target local check, and source support, then adds the items to semantic and structural indexes in a database-scoped repair memory.}
  \label{fig:offline-memory-items}
  \vspace{-2em}
\end{figure}

\subsubsection{Repair Unit Identification}
\label{sec:repair-unit-identification}

After Repair Recovery, the difference between \(s_i^{-}\) and \(s_i^{\mathrm{val}}\) reveals what changed, but not how those changes should be divided into repair units. Treating every AST edit as a separate unit can split one repair across multiple items. Treating the complete diff as one unit instead mixes distinct repairs. Syntax alone also cannot show whether a change is required by the historical task. Repair Unit Identification first combines dependent edits into structural groups. It then isolates one group at a time while keeping all other changes fixed. A group is accepted only when the historical task and database evidence show that it corrects a required behavior.

\stitle{Grouping dependent changes.}
Fine-grained tree differencing has long been used to extract structural source-code changes~\cite{fluri2007changedistiller,falleri2014differencing}. \method{} parses \(s_i^{-}\) and \(s_i^{\mathrm{val}}\) and derives neutral \texttt{ADD}, \texttt{REMOVE}, and \texttt{REPLACE} operations from their abstract syntax trees. Each operation records its clause, before-and-after fragments, and referenced schema elements. Equivalent changes and surface-only alias changes are discarded. The remaining edits are grouped by structural dependencies. For example, a join that introduces a table is grouped with edits whose column references require that table. Repeated instances of the same structural transformation also stay in one group. This step identifies edits that must be judged together without yet assigning them corrective meaning.

\stitle{Isolating one group.}
For each structural group \(g\), \method{} reverts only \(g\) from \(s_i^{\mathrm{val}}\) and leaves all other changes in place, producing \(s_i^{\neg g}\). It reparses the two queries to confirm that \(g\) is their only structural difference; otherwise, the group is discarded. This comparison isolates what \(g\) changes while preventing the other repairs from affecting the judgment.

\stitle{Judging the group from evidence.}
For every valid pair, \method{} collects parser-derived properties, schema and column profiles, query-grain statistics, relevant predicate or join observations grounded in the database content~\cite{brunner2021valuenet}, and the execution contrast between \(s_i^{\neg g}\) and \(s_i^{\mathrm{val}}\). A semantic judge derives the required behavior from \(q_i\) and determines whether the two SQL forms differ on that behavior using the supplied facts. A group is not accepted merely because it appears in the validated repair SQL or changes the execution result. If a decisive database fact is missing, the judge may request a targeted read-only probe before making its final decision.

A group is accepted as a repair unit only when \(q_i\) and objective facts support that \(s_i^{\mathrm{val}}\) implements the required behavior while \(s_i^{\neg g}\) conflicts with it. A repair unit may therefore contain multiple SQL edits. In Figure~\ref{fig:offline-memory-items}, adding \texttt{DISTINCT} and correcting the region literal binding form separate repair units. Each addresses a distinct requirement and is evaluated while the other repair remains fixed.

\subsubsection{Memory Item Construction}
\label{sec:memory-item-construction}

An accepted repair unit is still tied to its source task. Reusing it later requires explicit conditions for retrieval, activation, and adaptation. Memory Item Construction therefore converts each accepted repair unit into one memory item, as illustrated in Figure~\ref{fig:offline-memory-items}. Repair units from the same historical correction remain separate.

\stitle{Memory item representation.}
Each memory item contains four components. Its \emph{semantic contract} records its applicability, the incorrect and required behaviors, the repair behavior, and preservation constraints. Its \emph{structural signature} records the edit operations, affected clauses and fragments, schema references, and minimum context needed to recognize the incorrect form. Its \emph{target local check} specifies the database observation to inspect and the violation signal that indicates recurrence. Its \emph{source support} retains the historical requirement, the \(s_i^{-}\)-to-\(s_i^{\mathrm{val}}\) comparison, and the objective facts used to accept the repair unit. Only the first three components are exposed during online reuse. Source support records provenance rather than providing a case to copy.

\stitle{Admission and indexing.}
A memory item is stored only when its structural transformation is complete and its semantic contract specifies both applicability and required behavior. \method{} encodes the applicability condition and required behavior as a semantic vector. It derives structural keys from the incorrect SQL form, referenced schema elements, and minimum context. For an \texttt{ADD} repair, the missing structure cannot appear in the incorrect SQL, so the keys instead describe its insertion context. The resulting memory items form a fixed, database-scoped repair memory before online reuse.

\subsection{Online Stage: Repair Memory Reuse and Adaptation}
\label{sec:repair-memory-reuse}

Given a current task \(q\), current SQL \(s^{\mathrm{cur}}\), database \(\mathcal D\), and the corresponding repair memory, Repair Memory Reuse and Adaptation returns an enhanced SQL \(\widetilde{s}\). A retrieved memory item is only a candidate for verification: retrieval does not show that its recorded error recurs, and its source repair does not specify how to edit the current SQL. The online stage therefore retrieves candidate items, activates only those supported by the current task, SQL, and database observations, and adapts the activated items while preserving unrelated logic. It returns a rewrite only after mechanical validation; otherwise, it retains \(s^{\mathrm{cur}}\).

\subsubsection{Memory Item Retrieval}
\label{sec:memory-item-retrieval}

Retrieval-augmented Text-to-SQL commonly uses similar demonstrations~\cite{zhang2023refsql}. Repair-memory retrieval, however, must cover two ways in which a repair may recur. Tasks with similar requirements may express them through different SQL structures, while the same incorrect structure may appear under different questions. Memory Item Retrieval therefore searches the repair memory of the current database through complementary semantic and structural channels.

\stitle{Semantic and structural retrieval.}
The semantic channel compares an embedding of \(q\) with each item's applicability and required behavior. The structural channel parses \(s^{\mathrm{cur}}\) and counts matches between its AST-derived keys and the item's structural signature. For memory item \(m\), the two scores are
\[
  \begin{aligned}
  \operatorname{score}_{\mathrm{sem}}(m,q)
  &=
  \cos\!\left(
    \operatorname{Enc}(q),
    \operatorname{Enc}(\operatorname{app}(m)\mathbin{\|}\operatorname{req}(m))
  \right),\\
  \operatorname{score}_{\mathrm{str}}\!\left(m,s^{\mathrm{cur}}\right)
  &=
  \left|K(m)\cap K\!\left(s^{\mathrm{cur}}\right)\right|,
  \end{aligned}
\]
where \(K(m)\) describes the item's incorrect structure and minimum context, and \(K(s^{\mathrm{cur}})\) contains the corresponding keys from the current SQL. For an \texttt{ADD} repair, structural matching uses the tables, columns, and clauses surrounding the missing insertion point.

\stitle{Merging candidate rankings.}
\method{} merges the two rankings within the available context budget and removes duplicates, alternating between channels when not all same-database items fit. Each candidate includes its semantic contract, structural signature, target local check, and retrieval channel, but not the complete historical correction. Retrieval determines which items are examined next; it does not activate them.

\subsubsection{Evidence-Verified Memory Activation}
\label{sec:memory-activation}

Retrieval establishes relatedness, not recurrence. A candidate may match the current task or SQL structure even when its recorded error is absent. Evidence-Verified Memory Activation therefore tests two conditions: the current task must require the behavior recorded by the item, and the current SQL must exhibit its recorded violation. The item is activated only when both conditions hold.

For each candidate \(m\), the solver checks its semantic contract against \(q\), including its non-applicability conditions. It then checks whether the error behavior or violation signal recorded by \(m\) occurs in \(s^{\mathrm{cur}}\). Semantic similarity may indicate a relevant requirement, and structural overlap may identify a relevant SQL location, but neither substitutes for these checks.

\stitle{Current database observations.}
To test recurrence, \method{} provides the solver with the complete current SQL, database schema, and bounded observations from the current database. It automatically inspects selected structures already present in \(s^{\mathrm{cur}}\), including string predicates and scalar transformations. These observations can reveal whether a recorded violation occurs, but the required behavior must still come from \(q\).

If one database fact is still needed, the solver may request one bounded, read-only \texttt{SELECT}/\texttt{WITH} probe. \method{} executes the probe and returns its observation for the final \texttt{keep} or \texttt{rewrite} decision. The probe supplies a database fact; it cannot create a task requirement. In Figure~\ref{fig:method-overview}, \(m_j\) is activated because its requirement is relevant and its recorded violation is observed. Item \(m_k\) is ignored because these conditions are not met.

\subsubsection{Target Local Adaptation}
\label{sec:target-adaptation}

An activated memory item specifies a required behavior, not a target edit. The historical and current queries may realize that behavior through different SQL structures, and several activated items may affect the same clause. Target Local Adaptation must therefore instantiate each repair in the current SQL without transferring source-specific changes.

\stitle{Bind, rewrite, and preserve.}
The solver binds each activated item to the relevant tables, columns, values, and SQL locations using its semantic contract, structural signature, and current database observations. Preservation constraints identify logic that should remain unchanged. When several items are activated, the solver integrates them into one complete rewrite rather than applying their source edits independently. Figure~\ref{fig:method-overview} shows the activated item \(m_j\) being bound to the current SQL and adapted into a candidate while its preservation constraints are retained.

\stitle{Validate or retain.}
Before returning the candidate, \method{} checks that it is nonempty, differs from \(s^{\mathrm{cur}}\), and parses as one read-only \texttt{SELECT}/\texttt{WITH} query~\cite{scholak2021picard}. It must also execute successfully on \(\mathcal D\) and return at least one output column. A candidate with the same observed result as \(s^{\mathrm{cur}}\) is treated as a no-op. If any check fails, \method{} retains \(s^{\mathrm{cur}}\); otherwise, it returns the candidate as \(\widetilde{s}\).

\section{Evaluation}
\label{sec:evaluation}

Our evaluation addresses four questions. \textbf{RQ1} evaluates whether
\method{} improves SQL correctness across benchmarks and upstream Text-to-SQL
systems while avoiding regressions on correct current SQL. \textbf{RQ2}
measures the cost of Repair Memory Construction and Repair Memory Reuse and Adaptation.
\textbf{RQ3} uses paired ablations to isolate the granularity at which
historical corrections are retained, when retrieved experience should
influence the current task, and how supported experience should be adapted to
the current SQL. \textbf{RQ4} examines whether successful repairs arise from
individual memory items or from composing multiple memory items.

\subsection{Experimental Setup}
\label{sec:experimental-setup}

\stitle{Benchmarks and evaluation settings.}
We evaluate on the 498-query corrected BIRD~\cite{li2023bird} Mini-Dev collection
released by Jin et al.~\cite{jin2026pervasive} and the 299-query ScienceBenchmark
development set~\cite{zhang2023sciencebenchmark}, covering 11 and three
databases, respectively. For each benchmark, we use first-attempt SQL generated
by CHESS~\cite{talaei2024chess}, DeepEye-SQL~\cite{li2025deepeyesql}, and
OmniSQL-32B~\cite{li2025omnisql}, yielding six evaluation settings. Within
each database, we randomly assign approximately 25\% of the queries to the
training set and the remaining 75\% to the test set. Confirmed corrections in
the training set are used only to construct method-specific historical
information; \method{} does not update model parameters. Each BIRD setting
contains 127 training and 371 test
queries, while each ScienceBenchmark setting contains 75 training and 224 test
queries. Overall, the six settings contain 606 training queries and 1,785 test
queries.

Following Section~\ref{sec:sql-correction}, the query presented to a corrector
is the current SQL \(s^{\mathrm{cur}}\), and the query returned by a corrector
is the enhanced SQL \(\widetilde{s}\). Within each setting, all correction
methods receive the same train/test split and exactly the same current SQL. We
include every test query when computing EX, successful repairs, and
regressions; a query whose correctness cannot be established by execution is
counted as incorrect.

\stitle{Correction model and evaluation protocol.}
All stages that use a general-purpose correction LLM employ
\texttt{gpt-5-2025-08-07}. For SQLFixAgent and SHARE, we retain the specialized
local models provided by their released implementations. Given a current task
\((q,s^{\mathrm{cur}},\mathcal D)\), a corrector can access only the
natural-language question, benchmark-provided task evidence, the database
schema and data instance, the current SQL, and the historical information
permitted by the corresponding method. The gold SQL \(s^{*}\) is used only
after correction and is never included in a correction prompt. We determine
correctness using the official execution-based evaluator of each
benchmark~\cite{li2023bird,zhang2023sciencebenchmark}.

\stitle{Comparators.}
The four comparators represent different ways of incorporating corrective
information. MAGIC~\cite{askari2025magic} distills historical error traces into
global self-correction guidelines. TK-Boost~\cite{agarwal2026tkboost} uses
historical knowledge to provide iterative feedback at the CTE level.
SQLFixAgent~\cite{cen2025sqlfixagent} combines retrieval of similar repair
examples with multi-agent correction. SHARE~\cite{qu2025share} uses a trained
specialized model for action-level refinement. TK-Boost was originally
designed to refine CTEs inside an upstream Text-to-SQL system. Under our common
post-generation correction interface, we instead provide its released SDK
with the same current SQL as every other method and report this result as
TK-Boost (adapted).

\stitle{Metrics.}
Our primary metric is execution accuracy (EX). We additionally report the
successful repairs \(N_{\mathrm{fix}}\), regressions \(N_{\mathrm{reg}}\), and
net change \(\Delta\mathrm{EX}\) defined in
Section~\ref{sec:historical-experience-setting}. EX measures final
correctness, whereas the paired counts separate repaired initially incorrect queries
from losses caused by changing correct current SQL.

\subsection{Main Results}
\label{sec:overall-effectiveness}

\begin{table*}[t]
  \centering
  \caption{Execution accuracy (\%) across benchmarks and upstream
  Text-to-SQL systems. Overall gains are relative to the common current SQL.
  Fix/Reg.\ reports successful repairs and regressions over all 1,785 test
  cases. Best results are in bold.}
  \label{tab:main-results}
  \footnotesize
  \renewcommand{\arraystretch}{1.15}
  \begin{tabular*}{\textwidth}{@{\extracolsep{\fill}}lrrrrrrrr@{}}
    \toprule
    & \multicolumn{3}{c}{BIRD}
    & \multicolumn{3}{c}{ScienceBenchmark}
    & \multicolumn{1}{c}{Overall}
    & \multicolumn{1}{c}{Fix/Reg.} \\
    \cmidrule(lr){2-4}\cmidrule(lr){5-7}\cmidrule(lr){8-8}
    Method & CHESS & DeepEye & OmniSQL & CHESS & DeepEye & OmniSQL
      & EX (\(\Delta\) pp) &  \\
    \midrule
    Current SQL
      & 69.00 & 80.05 & 65.50 & 41.96 & 51.79 & 55.36
      & 63.31 (\(\text{--}\)) & \(\text{--}\) \\
    MAGIC
      & 77.09 & 86.25 & 84.10 & \textbf{57.14} & 54.91 & 57.14
      & 72.66 (\(+9.36\)) & 247/80 \\
    TK-Boost\textsuperscript{\(\dagger\)}
      & 73.85 & 76.82 & 71.70 & 50.45 & 55.80 & 57.14
      & 66.72 (\(+3.42\)) & 239/178 \\
    SQLFixAgent
      & 75.20 & 80.86 & 74.93 & 44.64 & 50.45 & 50.00
      & 66.22 (\(+2.91\)) & 114/62 \\
    SHARE
      & 80.59 & 83.02 & 83.56 & 45.09 & 50.45 & 49.11
      & 69.52 (\(+6.22\)) & 228/117 \\
    \method{}
      & \textbf{86.25} & \textbf{90.57} & \textbf{87.33}
      & 53.13 & \textbf{61.61} & \textbf{60.71}
      & \textbf{76.92 (\(+13.61\))} & \textbf{261/18} \\
    \bottomrule
  \end{tabular*}
  \vspace{2pt}

  \parbox{0.97\textwidth}{\scriptsize
  OmniSQL abbreviates OmniSQL-32B.
  \textsuperscript{\(\dagger\)}TK-Boost originally performs iterative CTE-level
  refinement inside an upstream Text-to-SQL system; here it is evaluated under
  the common post-generation correction interface.}
\end{table*}

\stitle{\method{} achieves the highest overall execution accuracy.}
As shown in Table~\ref{tab:main-results}, the current SQL is correct on 1,130
of the 1,785 test cases, corresponding to 63.31\% EX. The enhanced SQL
produced by \method{} is correct on 1,373 cases, yielding 76.92\% EX. This is a
net gain of 243 correct cases, or 13.61 percentage points. Among the 655
initially incorrect queries, \method{} successfully repairs 261. Among the
1,130 initially correct queries, it introduces only 18 regressions. The
improvement therefore combines substantial repair coverage with conservative
intervention on already-correct current SQL.

\begin{figure}[t]
  \centering
  \includegraphics[width=.88\columnwidth]{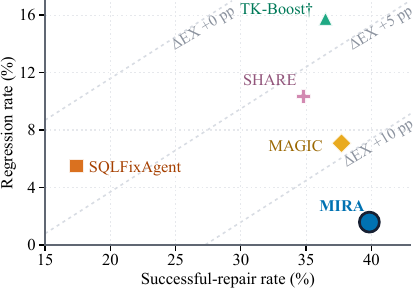}
  \caption{\method{} achieves the strongest repair--regression trade-off.
  Rates are computed over 655 initially incorrect and 1,130 initially correct
  queries, respectively; lower-right is better. Dotted lines connect equal net
  EX gains, and TK-Boost denotes the adapted implementation.}
  \Description{A scatter plot comparing five correction methods. \method{} is
  farthest toward the lower-right, combining the highest successful-repair
  rate with the lowest regression rate.}
  \label{fig:repair-regression-tradeoff}
  \vspace{-2em}
\end{figure}

\stitle{The main advantage is a better balance between successful repairs
and regressions.}
\method{} repairs 39.85\% of the initially incorrect queries and regresses
only 1.59\% of the initially correct queries. MAGIC, the strongest comparator
by overall EX, obtains a 37.71\% successful-repair rate and a 7.08\%
regression rate. Although the two methods differ by only 14 successful
repairs, \method{} produces 62 fewer regressions. It consequently retains 76
more correct cases in total, corresponding to a 4.26-point EX advantage.
TK-Boost (adapted) and SHARE repair 239 and 228 incorrect queries,
respectively, but also regress 178 and 117 correct queries. SQLFixAgent causes
fewer regressions, but repairs only 114 incorrect queries.
Table~\ref{tab:main-results} reports exact EX and paired outcomes;
Figure~\ref{fig:repair-regression-tradeoff} isolates the resulting trade-off.

\stitle{The improvements generalize across benchmarks and upstream
systems.}
\method{} improves the current SQL in all six combinations of benchmarks and
upstream systems and achieves the highest EX in five. On BIRD, aggregate EX increases
from 71.52\% to 88.05\%, a gain of 16.53 points produced by 192 successful
repairs and only 8 regressions. On ScienceBenchmark, aggregate EX increases
from 49.70\% to 58.48\%, a gain of 8.78 points produced by 69 successful
repairs and 10 regressions. Grouped by upstream system, the gains over CHESS,
DeepEye, and OmniSQL-32B are 14.96, 10.25, and 15.63 points, respectively.
The benefit therefore does not depend on a particular upstream system or its
distribution of current SQL.

\stitle{Generalization to ScienceBenchmark.}
On ScienceBenchmark, SQLFixAgent and SHARE, both of which rely on trained
specialized components, reduce EX by 1.34 and 1.49 points, respectively. By
contrast, MAGIC, TK-Boost (adapted), and \method{} incorporate historical
corrective information without updating the corrector parameters for the
target benchmark and improve EX by 6.70, 4.76, and 8.78 points. \method{}
makes 69 successful repairs, slightly fewer than the 72 of MAGIC and the 82
of TK-Boost, but introduces only 10 regressions, compared with 27 and 50.
This lower regression count gives \method{} the highest aggregate EX on
ScienceBenchmark. Together with its 16.53-point gain on BIRD, the results
show consistent transfer across databases, benchmarks, and upstream systems
while limiting unnecessary intervention under distribution change. The only
setting in which \method{} is not best is ScienceBenchmark--CHESS, where
MAGIC obtains 57.14\% and \method{} obtains 53.13\%.

\subsection{Efficiency}
\label{sec:efficiency}

Avoiding task-specific parameter updates does not eliminate the cost of memory
construction and inference. Table~\ref{tab:efficiency}(a) therefore reports
final EX and model usage over all 1,785 inputs. For SQLFixAgent and SHARE,
calls and tokens include both their specialized local models and the
general-purpose correction LLM. For the remaining methods, the measurements
cover their complete pipelines under our common protocol.

\begin{table}[t]
  \centering
  \caption{Model usage. Panel (a) compares complete correction pipelines over
  all 1,785 inputs. Panel (b) separates the one-time offline construction cost
  of \method{} from its online enhancement cost. Token counts are in millions.}
  \label{tab:efficiency}
  \footnotesize
  \renewcommand{\arraystretch}{1.12}

  \textbf{(a) Complete correction pipelines}\\[2pt]
  \begin{tabular*}{\columnwidth}{@{\extracolsep{\fill}}lrrr@{}}
    \toprule
    Method & EX (\%) & Calls & In/Out tok. (M) \\
    \midrule
    MAGIC & 72.66 & 2,365 & 22.03/4.96 \\
    TK-Boost\textsuperscript{\(\dagger\)} & 66.72 & 31,654 & 144.87/57.94 \\
    SQLFixAgent & 66.22 & 5,881 & 5.54/3.18 \\
    SHARE & 69.52 & 8,902 & 5.86/2.45 \\
    \method{} & \textbf{76.92} & 4,779 & 13.71/6.12 \\
    \bottomrule
  \end{tabular*}

  \vspace{5pt}
  \textbf{(b) Offline and online usage of \method{}}\\[2pt]
  \begin{tabular*}{\columnwidth}{@{\extracolsep{\fill}}lrrr@{}}
    \toprule
    Phase & Logical & Chat/Emb. & In/Out tok. (M) \\
    \midrule
    Offline construction & 1,174 & 1,132/42 & 2.73/3.02 \\
    Online enhancement & 3,598 & 1,820/1,778 & 10.99/3.10 \\
    \bottomrule
  \end{tabular*}
  \vspace{-1em}
\end{table}

\stitle{\method{} attains the highest EX without requiring the most model
computation.}
Relative to TK-Boost (adapted), \method{} uses 84.9\% fewer model calls and
90.2\% fewer total tokens while increasing EX by 10.20 points. Compared with
MAGIC, \method{} makes approximately twice as many calls but uses 26.5\% fewer
total tokens and improves EX by 4.26 points. SQLFixAgent and SHARE consume
fewer tokens, but their EX is lower by 10.70 and 7.39 points, respectively.
\method{} also makes 18.7\% fewer calls than SQLFixAgent and 46.3\% fewer
calls than SHARE. Thus, the repair--regression advantage in
Table~\ref{tab:main-results} does not depend on the exceptionally large
test-time computation used by TK-Boost, while the lower-token alternatives
provide substantially weaker correction accuracy.

\stitle{Offline construction is amortized, while online enhancement
requires about two logical requests per query.}
Table~\ref{tab:efficiency}(b) separates the one-time cost of building the
database-scoped memory item store from the online cost incurred for each test
question. Offline construction uses 1,174 logical requests and accounts for
24.6\% of all logical requests. It can be amortized across subsequent queries,
and incorporating a new confirmed historical correction does not update the
corrector parameters. The online stage uses 3,598 logical requests,
averaging 2.02 per input: approximately one retrieval embedding and 1.02 chat
requests. It consumes an average of 6,155 input tokens and 1,734 output tokens
per case. Panel (a) counts actual request attempts, whereas panel (b) reports
deduplicated logical requests.
\FloatBarrier

\subsection{Ablation Study}
\label{sec:ablations}
We conduct paired ablations on the 371 BIRD--DeepEye test cases, with the
full \method{} configuration (Full) as the reference. All variants use the same
inputs, GPT-5 configuration, context budget, and unchanged pipeline components.
Table~\ref{tab:ablations} reports results only on these 371 cases.

The variants remove one component at a time. \textbf{w/o Repair Unit
Decomposition} stores each historical incorrect--corrected SQL pair as a
single record. \textbf{w/o Evidence-Verified Memory Activation} uses an LLM
to select retrieved items from similarity-based context, without semantic
contracts or target-database verification. \textbf{w/o Target Local
Adaptation} passes the items activated by Full directly to the rewriter,
without mapping their repair requirements to the current SQL or specifying
what to modify and preserve.

\begin{table}[t]
  \centering
  \caption{Paired component ablations on 371 BIRD--DeepEye test cases.
  \(\Delta\) Full is the EX difference from the unchanged Full configuration.}
  \label{tab:ablations}
  \footnotesize
  \renewcommand{\arraystretch}{1.12}
  \begin{tabular*}{\columnwidth}{@{\extracolsep{\fill}}lrrr@{}}
    \toprule
    Setting & Fix/Reg. & EX (\%) & \(\Delta\) Full (pp) \\
    \midrule
    Current SQL & 0/0 & 80.05 & \(-10.51\) \\
    \textbf{Full} & \textbf{42/3} & \textbf{90.57} & \(\text{--}\) \\
    w/o Repair Unit Decomp. & 29/10 & 85.18 & \(-5.39\) \\
    w/o Verified Memory Act. & 40/14 & 87.06 & \(-3.50\) \\
    w/o Local Adaptation & 34/12 & 85.98 & \(-4.58\) \\
    \bottomrule
  \end{tabular*}
  \vspace{-1em}
\end{table}

\stitle{The components have different effects.}
The current SQL is correct in 297 of 371 cases (80.05\% EX). Full repairs 42
queries with 3 regressions, reaching 90.57\% EX. Removing decomposition mainly
reduces repairs, removing verified activation increases regressions, and
removing local adaptation affects both.

\stitle{Repair Unit Decomposition improves repair coverage.}
Using complete SQL pairs reduces EX from 90.57\% to 85.18\%: repairs fall from
42 to 29, while regressions rise from 3 to 10. A historical correction may
contain several changes, but the current SQL may require only some of them.
Decomposition separates these changes so that each can be evaluated and
reused independently.

\stitle{Evidence-Verified Memory Activation prevents unnecessary changes.}
The similarity-only variant retains 40 of Full's 42 repairs, but regressions
increase from 3 to 14. Thus, retrieval preserves most repair ability, whereas
semantic contracts and target-database verification help reject irrelevant
experience and protect correct SQL.

\stitle{Target Local Adaptation applies repairs to the current SQL.}
Without structured adaptation, repairs fall from 42 to 34 and regressions rise
from 3 to 12, even when using the items and evidence selected by Full. Passing
relevant items as prompt context alone may miss required changes or alter
unrelated logic. Explicitly mapping each repair to the current SQL improves
both completeness and precision.

Overall, decomposition identifies reusable changes, verified activation
selects relevant changes, and local adaptation applies them to the current
SQL. The distinct repair and regression patterns show that all three
components contribute to reliable correction.
\FloatBarrier

\subsection{In-depth Analysis}
\label{sec:in-depth-analysis}

\stitle{Independent reuse and composition of memory items.}
We examine how memory items contribute to the 261 successful repairs produced
by \method{}. A single-item repair instantiates one memory item in the enhanced SQL,
whereas a multi-item repair jointly instantiates two or more.

\begin{figure}[t]
  \centering
  \includegraphics[width=.86\columnwidth]{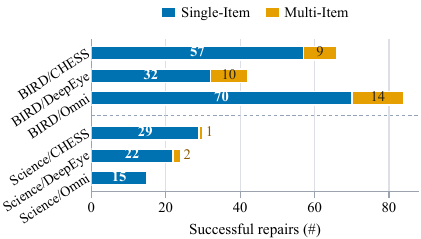}
  \caption{Memory items support independent reuse in all six
  benchmark--upstream-system settings and compositional repair in five. Each
  stacked bar reports successful single- and multi-item repairs. Science and
  Omni abbreviate ScienceBenchmark and OmniSQL-32B. Overall, 225 repairs use
  one item and 36 compose multiple items.}
  \Description{Six horizontal stacked bars divide successful repairs into
  single-item and multi-item repairs. Single-item repairs dominate every
  setting, while multi-item repairs occur in five settings.}
  \label{fig:item-repair-composition}
  \vspace{-2em}
\end{figure}

\stitle{Most memory items remain useful as independent repair units.}
Figure~\ref{fig:item-repair-composition} shows that 225 of the 261 successful
repairs are single-item repairs, accounting for 86.2\% of the total. They
occur in all six settings and constitute the majority in every setting, with
shares from 76.2\% to 100\%. In these cases, the framework isolates one
supported repair requirement from history and binds it to the corresponding
location in the current SQL. This is the intended effect of representing a
historical correction as fine-grained memory items: the system transfers the
change required by the current task rather than reusing the entire historical
correction.

\stitle{Multiple memory items can be composed in one current task.}
The remaining 36 repairs, or 13.8\%, require more than one memory item: 29
combine two items and 7 combine three. Multi-item repairs occur in five of the
six settings and span all three upstream systems. These cases require the
solver to identify several repair targets, bind them to distinct but
interdependent SQL locations, and realize them consistently in one enhanced
SQL. Their correctness under the official evaluators shows that independently
mined items can be recomposed into a coherent correction for a new task.
Together, the 225 single-item and 36 multi-item repairs demonstrate both
selective transfer and compositional repair.

\FloatBarrier

\FloatBarrier

\section{Related Work}
\label{sec:related-work}

\stitle{SQL Self-Correction.} Self-correction uses staged
reasoning~\cite{pourreza2023dinsql,shen2024selectsql,wang2025dac} or execution
and database feedback~\cite{wang2024toolsql,tian2026pvsql,shen2025maplerepair}
from the current task. \method{} instead reuses confirmed same-database
corrections.

\stitle{Training-Based SQL Correction.} Prior work learns correction and error
detection from training
data~\cite{chen2023sqlcorrection,qu2025share,hong2026errorllm,gong2025sqlens}.
SQLFixAgent combines a fine-tuned SQLTool with retrieved
repairs~\cite{cen2025sqlfixagent}. \method{} updates repair memory rather than
model parameters.

\stitle{Experience-Based SQL Correction.} Experience-based methods reuse prior
corrections at inference time. MAGIC distills histories into
guidelines~\cite{askari2025magic}; Memo-SQL retrieves complete
cases~\cite{yang2026memosql}; and TK-Boost applies structured knowledge inside
an NL2SQL agent through CTE refinement~\cite{agarwal2026tkboost}. \method{}
stores independent repair requirements as memory items for a separate
post-generation corrector. Each item excludes unrelated source changes and can
be selected, adapted, or combined independently.

\section{Conclusion}
\label{sec:conclusion}

We study training-free, experience-based correction of SQL
produced by upstream Text-to-SQL systems. \method{} converts confirmed
same-database corrections into independently reusable memory items, activates
items supported by the current question, SQL, and database evidence, and
adapts them while preserving unrelated logic. Across BIRD and
ScienceBenchmark with three upstream systems, \method{} improves EX by 16.53
and 8.78 percentage points, with 261 repairs and 18 regressions. These results
support storing history at independently judgeable repair boundaries and
applying it with evidence from the current task.

\bibliographystyle{ACM-Reference-Format}
\bibliography{references}

\end{document}